\documentclass[12pt]{article}

\usepackage{floatrow}
\usepackage{caption}

\usepackage{fullpage}
\usepackage{setspace}
\usepackage{parskip}
\usepackage{titlesec}
\usepackage[section]{placeins}
\usepackage{xcolor}
\usepackage{breakcites}
\usepackage{lineno}
\usepackage{hyphenat}
\usepackage{array}

\PassOptionsToPackage{hyphens}{url}
\usepackage[colorlinks = true,
            linkcolor = blue,
            urlcolor  = blue,
            citecolor = blue,
            anchorcolor = blue]{hyperref}
\usepackage{etoolbox}


\renewenvironment{abstract}
  {{\bfseries\noindent{\abstractname}\par\nobreak}\footnotesize}
  {\bigskip}

\titlespacing{\section}{0pt}{*3}{*1}
\titlespacing{\subsection}{0pt}{*2}{*0.5}
\titlespacing{\subsubsection}{0pt}{*1.5}{0pt}

\usepackage{authblk}

\usepackage{graphicx}
\usepackage{physics}
\usepackage[space]{grffile}
\usepackage{latexsym}
\usepackage{textcomp}
\usepackage{longtable}
\usepackage{tabulary}
\usepackage{booktabs,array,multirow}
\usepackage{amsfonts,amsmath,amssymb}
\providecommand\citet{\cite}
\providecommand\citep{\cite}

\newcommand\bea{\begin{eqnarray}}
\newcommand\eea{\end{eqnarray}}

\newif\iflatexml\latexmlfalse

\AtBeginDocument{\DeclareGraphicsExtensions{.pdf,.PDF,.eps,.EPS,.png,.PNG,.tif,.TIF,.jpg,.JPG,.jpeg,.JPEG}}

\usepackage[utf8]{inputenc}
\usepackage[english]{babel}

\newcommand{\Z}{\ensuremath{\mathbb Z}}

\begin{document}

%\title{Learning wavefunction of data: tensor network-based sampler-classifier. }
\title{Tensor network to learn the wavefunction of data. }

\author[1, 2]{Anatoly Dymarsky}%

\author[2]{Kirill Pavlenko}%

\affil[1]{Department of Physics and Astronomy, University of Kentucky, Lexington, Kentucky 40506, USA}%
\affil[2]{Skolkovo Institute of Science and Technology, Skolkovo Innovation Center, Moscow 143026, Russia}%

\vspace{-1em}

  \date{\today}

\begingroup
\let\center\flushleft
\let\endcenter\endflushleft
\maketitle
\endgroup

\selectlanguage{english}
\begin{abstract}
{How many different ways are there to handwrite digit 3? To quantify this question imagine extending  a dataset of handwritten digits MNIST by sampling additional images until they start repeating. We call the collection of all resulting images of digit 3  the ``full set.'' To study the properties of the full set we introduce a tensor network architecture which simultaneously accomplishes both classification (discrimination) and sampling tasks. Qualitatively, our trained network represents the indicator function of the full set.
It therefore can be used to characterize the data itself. We illustrate that by studying the  full sets associated with  the digits of MNIST.
 Using quantum mechanical interpretation of our network we characterize the full set by calculating its entanglement entropy. We also study its geometric properties such as mean Hamming distance, effective dimension, and size. The latter  answers the question above  -- the total number of black and white threes written MNIST style is $2^{72}$.}\\%
\end{abstract}%

\section*{Introduction}
Generalization is a remarkable ability of supervised learning algorithms to learn patterns underlying training data and subsequently perform well on new datasets. It reflects both potency of the algorithm, but also certain simplicity of the training data.  Namely presence of  patterns that might be apparent to a human eye but usually very difficult to quantify. On the contrary datasets without underlying patterns, such as fully random or ad hoc ones can be learned but can not be generalized \cite{zhang2017understanding,zhang2021understanding}. To better understand when generalization is possible and inform development of more efficient supervised learning algorithms, it would be important to  characterize  patterns that underlie various datasets of interest. In this context a training set should be thought of as a small subset of the ``full set of data’’ which includes all possible hypothetical data exhibiting given patterns. We introduce a novel tool, a tensor network sampler-discriminator/classifier, which learns the ``wave-function of data.”  Qualitatively, our tensor  network is  the indicator function of the full set and provides new ways to quantitatively study and characterize it. 

%For this task only MNIST images of “3” are used [may be used?] as the training data, as in \cite{}. 

To keep the presentation simple in what follows we focus on a particular example of MNIST, the dataset of handwritten digits. All ideas and techniques can be immediately extended to other instances of supervised learning. MNIST contains images measuring 28 by 28 pixels; we transform them from grayscale to black and white for simplicity, such that there are $2^{784}$ possible images in total. A standard task would be to train a classifier to distinguish different digits. To further simplify things, we can train the network to recognize  a particular digit, say digit 3, by distinguishing it from images of other digits, other symbols or noise. 
Contemporary architectures can achieve this task with a small generalization error, which implies that among all possible $2^{784}$ images one can define the set of all images of digit 3, which our discriminator network would recognize. Good quality of generalization exhibited by various machine learning architectures \cite{neyshabur2017exploring, kawaguchi2020generalization} suggests this full set can be defined with a large degree of objectivity, essentially in the architecture-independent way. 
%In this paper we outline the task of studying basic properties of the full set, including its geometric structure and complexity. %, as they control and may help improve effectiveness of the machine learning algorithms. 

Practically the full set is unfathomably large and is never available. In this work we propose a way to study it using a  tensor network, which mathematically  is an $L_2$-normalized ``wave-function'' $\Psi(x)$, deffined on the space $I=\Z_2^{784}$ of all $N_I=2^{784}$ possible images. Qualitatively function 
\bea
{\cal P}(x):=\left\{ \begin{array}{cc}
1, &\,{\rm for}\, \, \, |\Psi(x)|^2 \geq  \epsilon,\\
0, &{\rm for}\, \, |\Psi(x)|^2 < \epsilon,  
\end{array}\right.  \label{P} 
\eea
 with some appropriate $\epsilon$ is the indicator function of the full set.
To emphasize that $\Psi$ characterizes the data itself and its properties exhibit robust independence of the tensor network architecture we call it the wavefunction of {\it data}. Using quantum mechanical interpretation of $\Psi(x)$ we can characterize the full set by calculating its entanglement entropy. We also study geometric properties of the full set such as mean Hamming distance, effective dimension, and the size. 
The latter is simply the approximate total number of images recognized by our network as depicting the given digit. In contrast to the first two properties, which can be studied using training set alone, size is the global property of the full set.

\begin{figure}
\centering
\begin{minipage}{.49\textwidth}
  \centering
  \includegraphics[width=.7\linewidth]{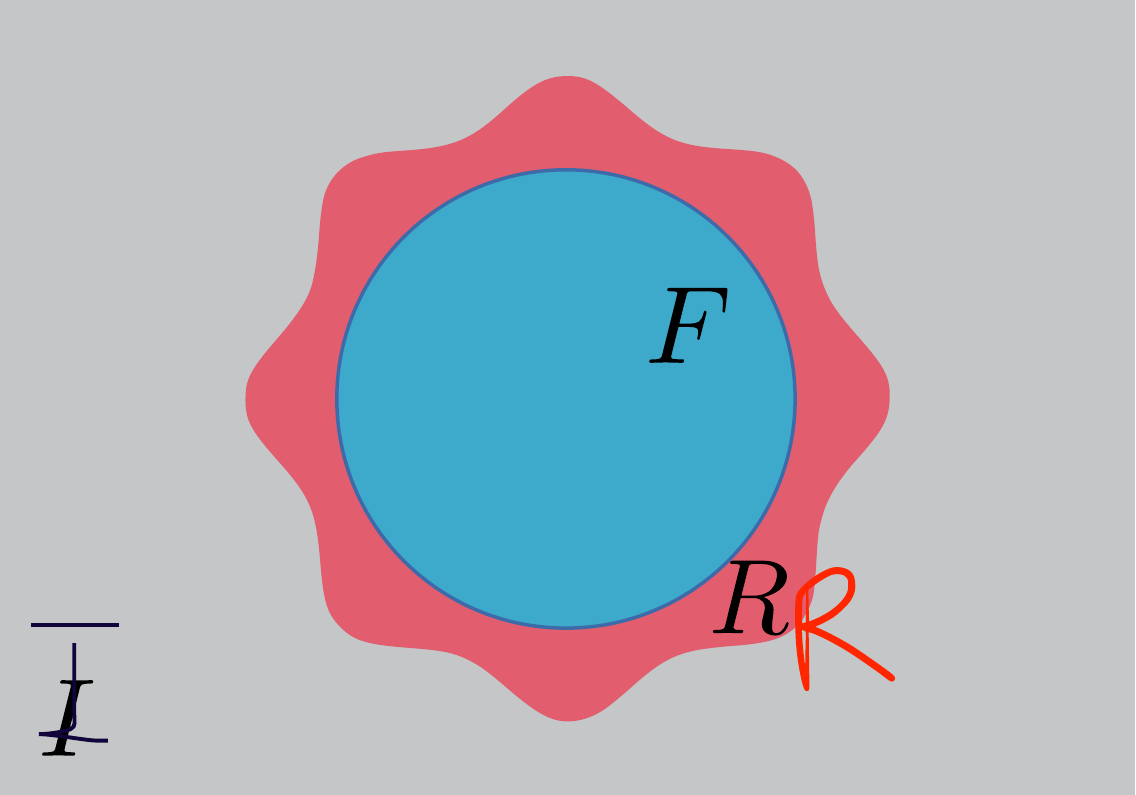}
\end{minipage}
\begin{minipage}{.49\textwidth}
  \centering
  \includegraphics[width=.7\linewidth]{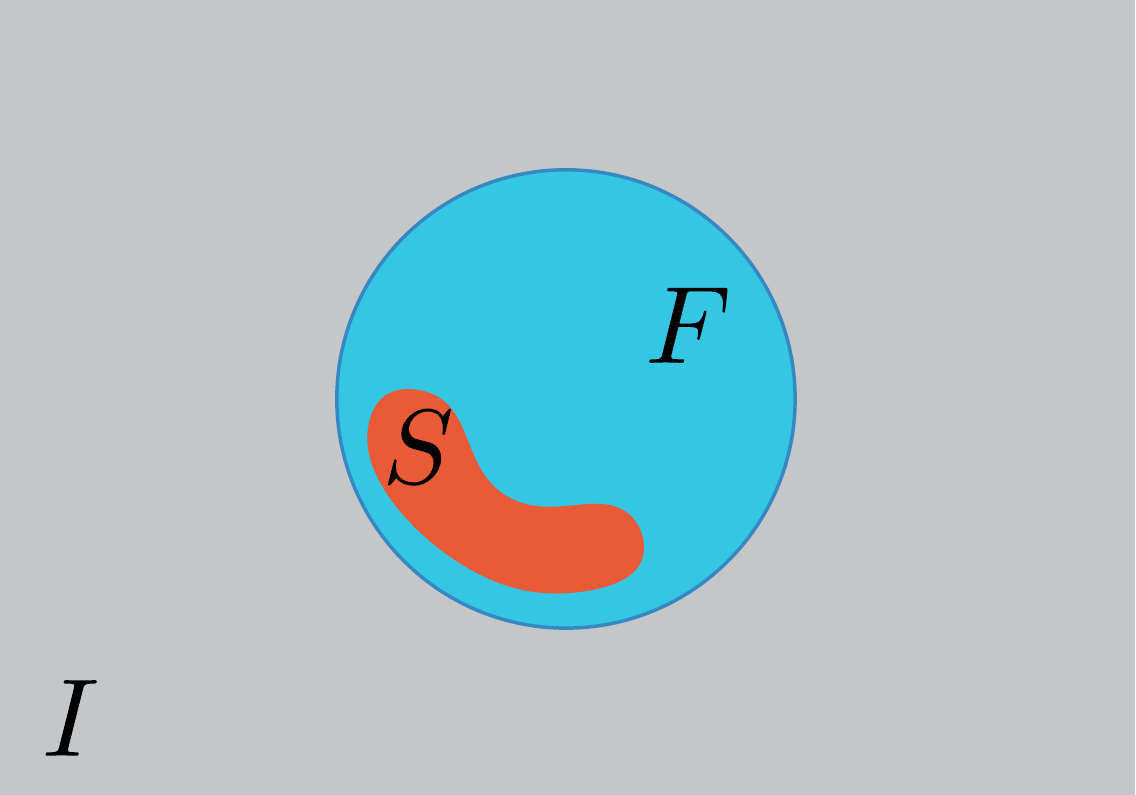}
\end{minipage}
\caption{
On both panels $I$ is the set of all $2^{784}$ images, $F$ is the full set of images of digit $i$.
Let panel: ideal discriminator $\Psi$ recognizes all images of digit $i$, but may also recognize as $i$ images of other digits or noise. 
This means $|\Psi(x)^2|\geq \epsilon$ for all $x\in F$, as well  as  for some set $x\in R$ depicted in red.
Right panel: all images sampled by an ideal sampler $\Psi$ are images of $i$. This means $\Psi$  has the support on the subset of the full set $S\subset F$, 
while 
$|\Psi(x)|^2 \approx 0$ for $x\notin S$. }
  \label{fig:discriminator}
\end{figure}

Before we proceed with the results, we would like to explain why tensor network
 sampler-discriminator/classifier is an appropriate architecture to define the full set via \eqref{P}.  In the recent years tensor networks,  such as Matrix Product States (MPS)  and Tensor Trains,  have been actively used  to build various classification  \cite{stoudenmire2016, novikov2017exponential,  efthymiou2019tensornetwork,  anagiannis2021entangled} and generative \cite{Han18generative,  Cheng19generative} algorithms.  They demonstrate robust performance on par with the advanced CNN architectures \cite{efthymiou2019tensornetwork, glasser2019expressive}. 
%From the mathematical point of view a trained tensor network defines a set of functions $\Psi_i(x)$ where $x$ represents the input data and $i$ stands for the labels. In the case of b $\&$ w MNIST $x$ is a vector of $28^2=784$ binary components $x_\alpha\in \{0,1\}$ and $i$ labels digits, $i=0\dots 9$. Then $p_i=|\Psi_i(x)|^2/\sum_j |\Psi_j(x)|^2$ is the probability that the image $x$ depicts digit $i$. One can further simplify the architecture and train only function $\Psi(x)\equiv \Psi_i(x)$ for a particular $i$, say $i=3$. Then the value of $|\Psi(x)|^2$, if exceeding a particular threshold $\epsilon$, would imply that the image $x$ depicts digit $i$, or does not otherwise. [Calibration of threshold.] 
In our case we train $\Psi$ for a particular digit $i$. Then the value ${\cal P}(x)=1$ means $\Psi$ recognizes $x$  as an image of $i$.\footnote{For the  classification task one needs to train ten networks $\Psi_i(x)$ for each $i$. Classification  is then performed by maximizing $p_i=|\Psi_i(x)|^2/\sum_j |\Psi_j(x)|^2$ over $i$, see Methods.}
Good quality of generalization  means our  network reliably recognizes images  $i$ outside of the training set. In the ideal case with perfect  quality of generalization there still could be images of other digits or even noise recognized by our network as $i$. This is illustrated in  the left panel of Fig.~\ref{fig:discriminator}. There gray square region represents all possible $N_I=2^{784}$ images. Red area $R$ represents images 
which our network ``recognizes'' as $i$,  i.e.~$|\Psi(x)|^2$  exceed $\epsilon$ for $x$ from this area. The blue disk, denoted as $F$  represents  the full set of images depicting  given digit $i$. It is  a subset of the red area $R$.

Tensor network architectures allow for an efficient evaluation of $|\Psi(x)|^2$ as a function of some components $x^\alpha$ while values of other components $x^\beta$ are fixed. It therefore can be used for sampling:  pixels  are sampled consequently,  using conditional probability distribution specified by $\Psi(x)$. This idea got traction recently and several such architectures were introduced in \cite{Han18generative,  Cheng19generative}. Clearly, only images with large values of $|\Psi(x)|^2$ can be sampled.  
%Quality of sampling can be accessed visually. To quantify it one can also use one of the classifier algorithms, e.g.~tensor models referenced above. We train an auxiliary neural network of \cite{}, which can gauge quality of sampling by assigning a probably for a given image of being  the image of digit $i$. 
Provided our sampler $\Psi(x)$ achieves a good quality, i.e.~ideally all sampled images depict $i$, we can think of $\Psi(x)$ as a function with the support on a subset of the full set. This is illustrated in the right panel of  Fig.~\ref{fig:discriminator}. There orange subset $S$ of the blue disk  represent  images $x$ for which  $|\Psi(x)|^2$ is sufficiently large to be sampled, while for all other $x\notin S$,  $|\Psi(x)|^2\approx 0$.

%The gray rectangular and blue disk have the same interpretation as in the left panel of  Fig.~\ref{fig:discriminator}, while the subset of the blue disk shown in orange stands for images $x$ for which  $|\Psi(x)|^2$ is sufficiently large to be sampled.

The idea of the sampler-discriminator is to  train an MPS-based tensor network $\Psi(x)$ which accomplishes both tasks. Schematically we minimize the objective function 
\bea
L=-{1\over N_T}\sum_{x\in T} \ln |\Psi(x)|^2 \label{lossf}
\eea
 where $T$ represents the training set -- a set of $N_T$ images of digit $i$. It is a small subset  inside the full set $F$.
%, which is presumably uniformly distributed within $F$.   
%blue disk from the figures  \ref{fig:classifier} and \ref{fig:sampler}. 
 Importantly, our architecture enforces  wave-function  normalization
\begin{eqnarray}
\sum_{x\in I} |\Psi(x)|^2=1, \label{normalization}\qquad I=\Z_2^{784}. 
\end{eqnarray}
As a result decreasing of the loss function \eqref{lossf} automatically decreases value of $|\Psi(x)|^2$ for $x$ outside of $T$. Assuming  $T$ is approximately uniformly distributed within $F$ and $\Psi(x)$ changes smoothly, we may expect $|\Psi(x)|^2$ to mostly decrease outside of $F$, while inside $F$ it would remain relatively large. The latter behavior would assure generalization of discriminator: value of $|\Psi(x)|^2$ for $x\in F$ would exceed certain threshold. The former property, smallness of $|\Psi(x)|^2$ for $x\notin F$, assures good quality of sampling. We thus conclude that a network $\Psi$ which simultaneously accomplishes both discrimination and sampling with high quality has a support on $F$, with \eqref{P} being its indicator function. 

In practice decreasing of the loss function during training process will eventually lead to overfitting when $|\Psi(x)|^2$ is large for $x\in T$ but not necessarily for $x\in F$. We therefore stop training as soon as discrimination/classification begins to reduce after reaching its maximal value. The logic outlined above is schematic, we justify it a posteriori by examining the quality of recognizing (classifying) and sampling achieved by the trained $\Psi$.  Further  details of the network architecture and the training process are described below in Methods. 

Ideally, for the trained network $\cal P$ defined in \eqref{P} is  the indicator function of the full set:  $|\Psi(x)|^2$  exceeds certain threshold for $x\in F$ and plunges below it for $x\notin F$. It therefore reflects the data itself rather than peculiarities of the architecture or the training process. To justify this claim we show that certain  properties of $\Psi(x)$,  such as quality of discriminating/ classifying and sampling,  typical value of  $|\Psi(x)|^2$ for $x\in F$,  value of entanglement entropy associated with $\Psi(x)$, etc.~are not sensitive to MPS bond dimension or initialization seed.  This confirms our main conclusion that the proposed architecture provides a novel way to quantitatively characterize the data itself, rather than peculiarities of the network design or the  training process.

\section*{Results}
The core of our construction is the Matrix Product State real tensor network in the canonical form \cite{Vidal2003}. Mathematically it is a real-valued function $\Psi(x)$ where $x^\alpha$ is a vector of  $28^2=784$ binary variables. Canonical form imposes normalization condition \eqref{normalization}. We train the network by minimizing loss function \eqref{lossf}
%\bea
%-{1\over N}\sum_{a=1}^N \ln |\Psi(x_\alpha)|^2, \label{lossf}
%\eea
via gradient descent, and the test set $T$ is the set of  black and white MNIST images of digit $i$. 
Corresponding tensor network is labeled $\Psi_i$. 
%Details of the architecture, training process, and performance tests are delegated to Methods.  

\begin{figure}
\centering
\begin{minipage}{.32\textwidth}
  \centering
\includegraphics[width=1\textwidth]{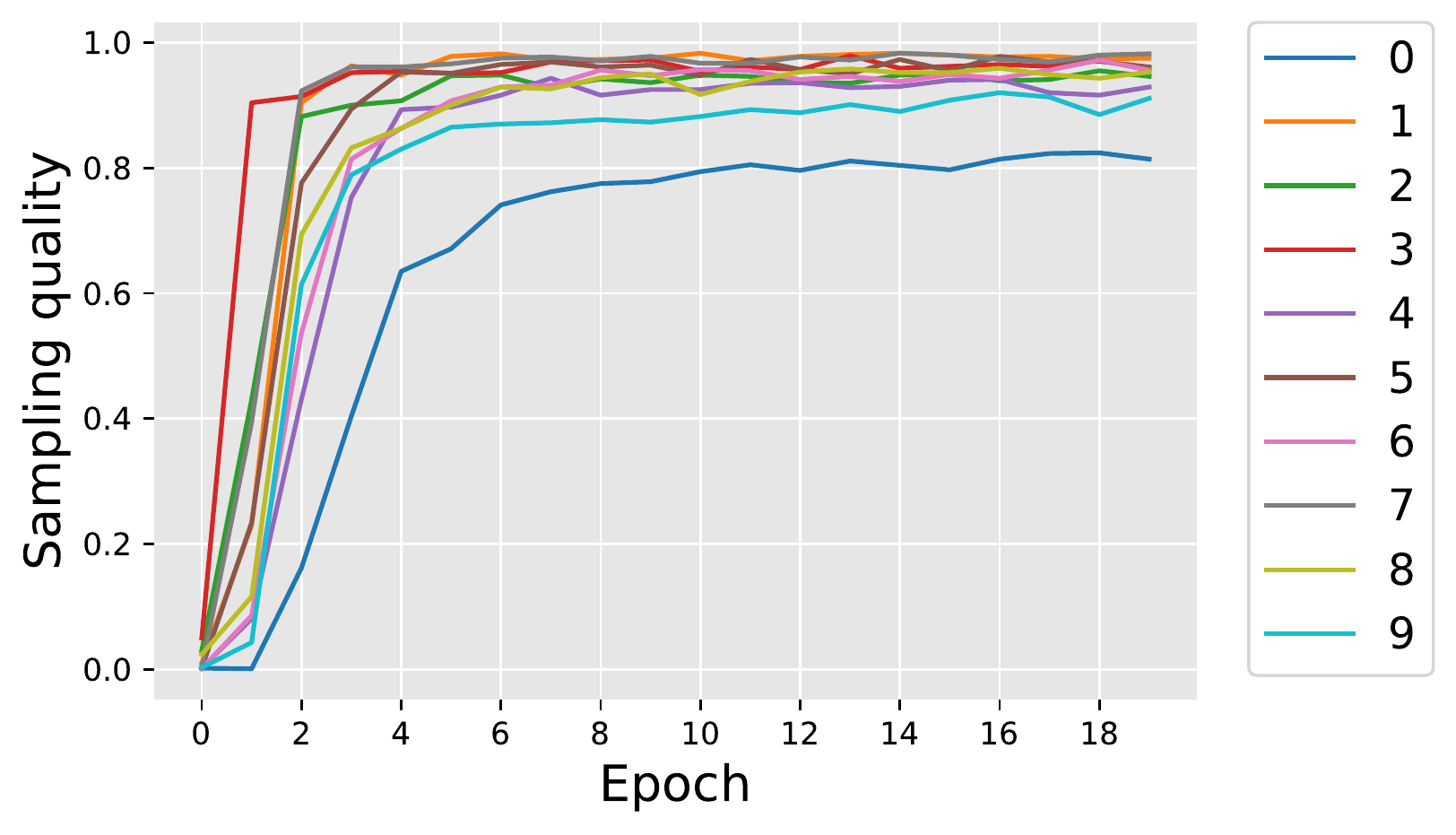}
\end{minipage}
\begin{minipage}{.265\textwidth}
  \centering
\includegraphics[width=1\textwidth]{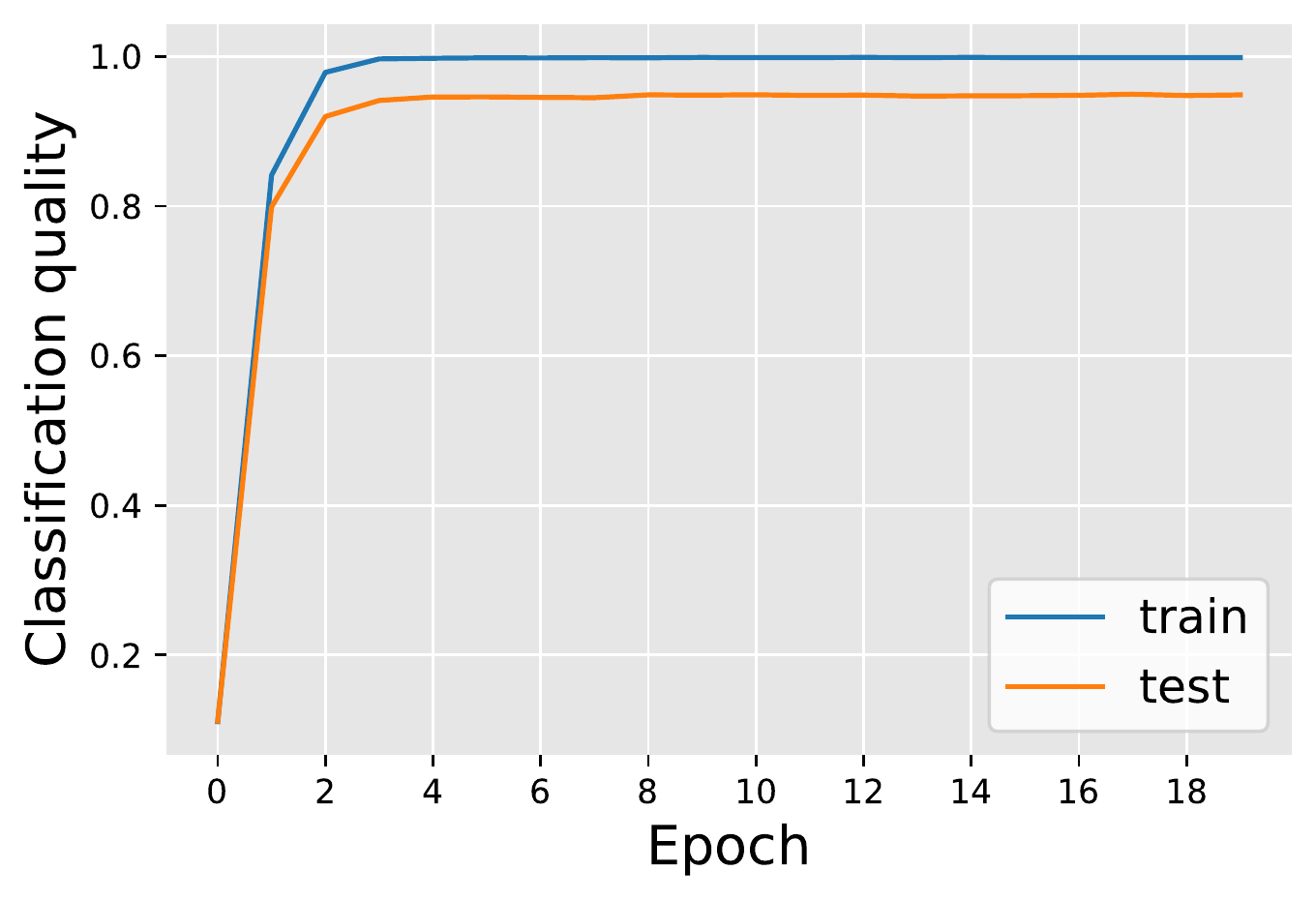}
\end{minipage}
\begin{minipage}{.265\textwidth}
  \centering
\includegraphics[width=1\textwidth]{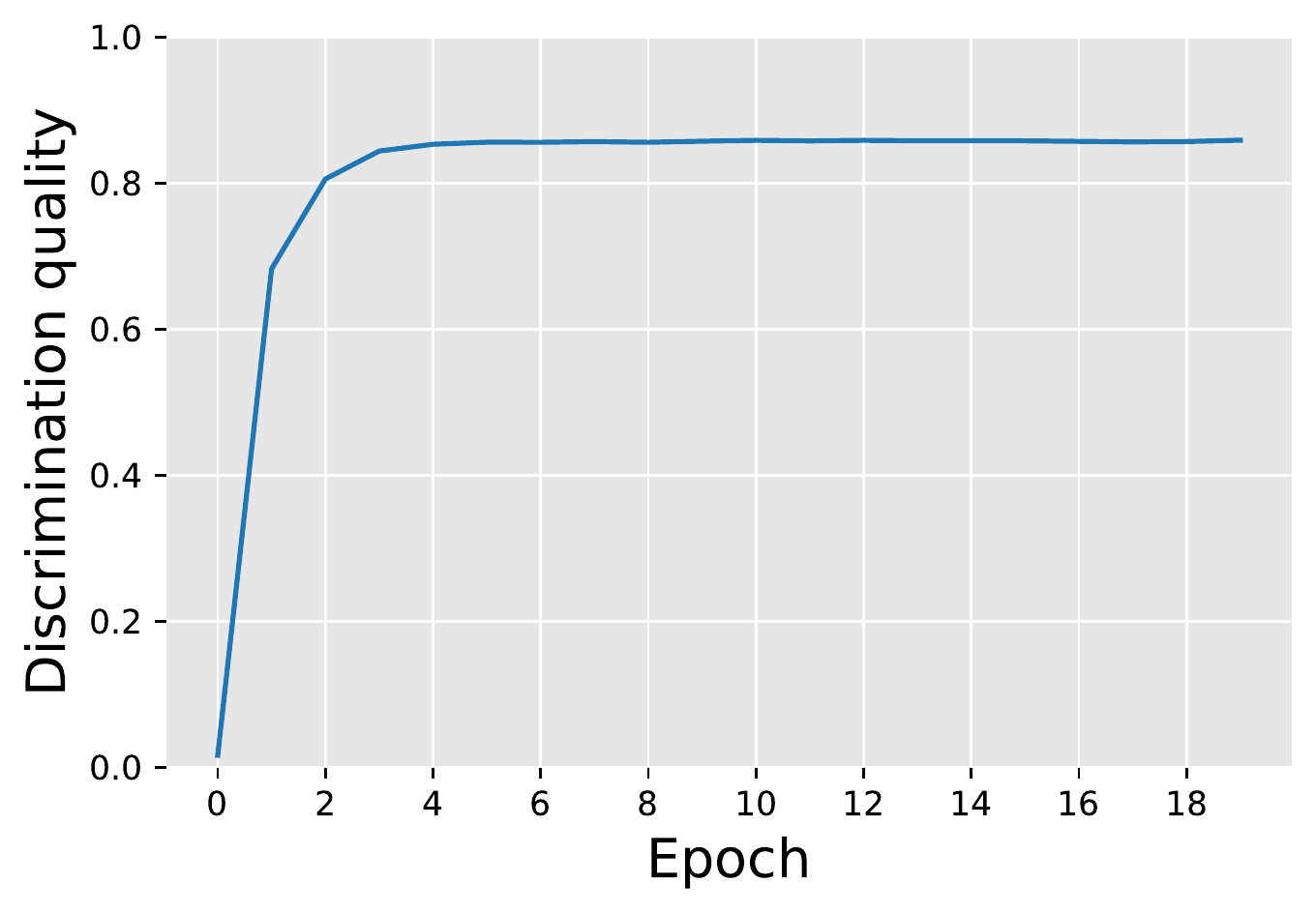}
\end{minipage}
\caption{Left panel: quality of sampling by $\Psi_3$ with the bond dimension $D = 100$ during the training process. Quality is assessed by an auxiliary CNN.  Central and right panels: quality of classification and  discrimination  by $\Psi_i$ and $\Psi_3$ with the same $D$ during the training process.}
\label{fig:training}
\end{figure}
%{Quality of sampling ($D = 50$) measured by CNN trained on QMNIST (extended MNIST). CNN was trained up to 99\% accuracy. Note, that quality of sampling of zeros is lower  than for other digits, as its entanglement is higher (see previous graph). This feature goes away with the increase of bond dim.}
%{This is score of \textbf{Classification} while training to \textbf{sample} by minimizing $- \frac{1}{N}\sum_{x \in training set} \log(P(x))$, where $P(x) = \frac{|\bra{x} \ket{\psi}|^2}{\bra{\psi} \ket{\psi}}$. The score is averaged over all digits (maybe it is better to separate them). Here bond dim $D = 50$. The graphs look similar for other bond dims with some signs of overfitting at large bonds.}

As the learning process proceeds, quality of sampling by $\Psi_i$ gradually grows -- the network remembers images from the training set and tries to replicate them. This is shown in the left panel of Fig.~\ref{fig:training}. The quality of recognizing digit $i$ for images from the test set 
(calibration of threshold $\epsilon$ is discussed in Methods)  grows initially, but then may decay slightly due to overfitting.  Similar behavior is exhibited by the quality of classification, for which all ten $\Psi_i$ must be trained.  This is shown in the right and central panels of Fig.~\ref{fig:training}. Overfitting becomes more pronounced when the bond dimension $D$ of the tensor network grows. To prepare the network of interest, which would simultaneously accomplish both sampling and discrimination/classification tasks,  the training process is stopped as soon as  the quality of discrimination/classification reaches its maximum. For the sufficiently large $D \gtrsim  100$ this happens already after a few epochs.
\begin{figure}[b]
\begin{minipage}{.49\textwidth}
  \centering
\includegraphics[width=0.9\textwidth]{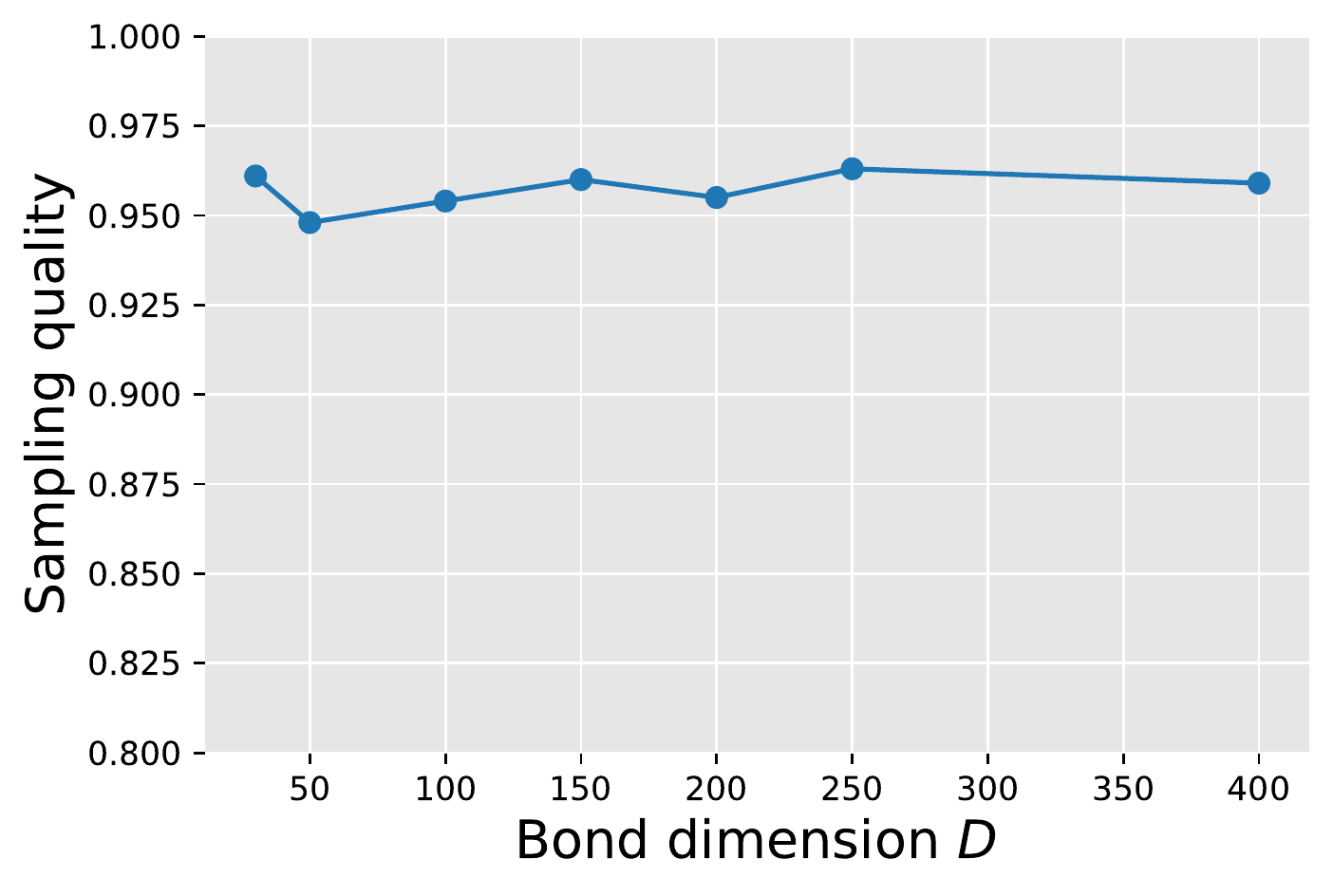}
\end{minipage}
\begin{minipage}{.49\textwidth}
\centering
\includegraphics[width=0.9\textwidth]{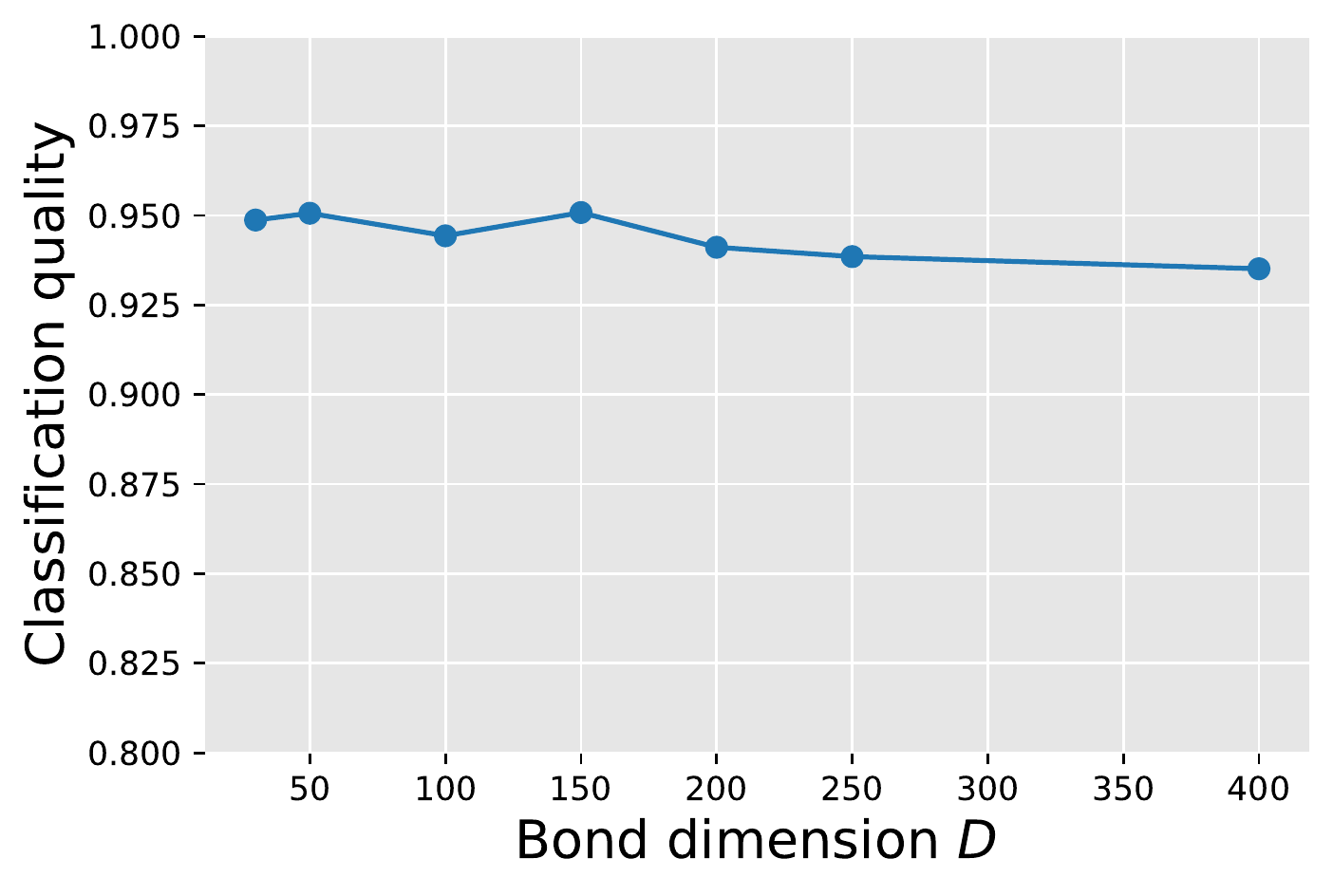}
\end{minipage}
\caption{Quality of sampling (left) and classification (right)  by the trained $\Psi_3$ as a function of bond dimension.  }
\label{fig:quality}
\end{figure}

We now demonstrate that core properties of properly trained $\Psi_i$ are largely independent of the  bond dimension $D$, provided the latter  is sufficiently large, $D \gtrsim 30$. 
To begin with we  study how the quality of sampling and classification depends on the bond dimension. The quality of  classification is the maximal value from the central panel of Fig.~\ref{fig:training}, since we stop training at that point. Results for sampling and classification for different $D$ shown in   Fig.~\ref{fig:quality} confirm that quality remains essentially the same in a wide range of bond dimensions. It is also not sensitive to the initial seed.

%Function \eqref{P} is then qualitatively the indicator function of the full set of images of digit $i$. In what follows we use $\Psi_i$ to study different properties of the full set and in parallel demonstrate their robust independence on $D$ and the initial seed.
%[robust independence of sampling quality etc. on D and the seed]

\begin{figure}
\begin{center}
\includegraphics[width=0.6\textwidth]{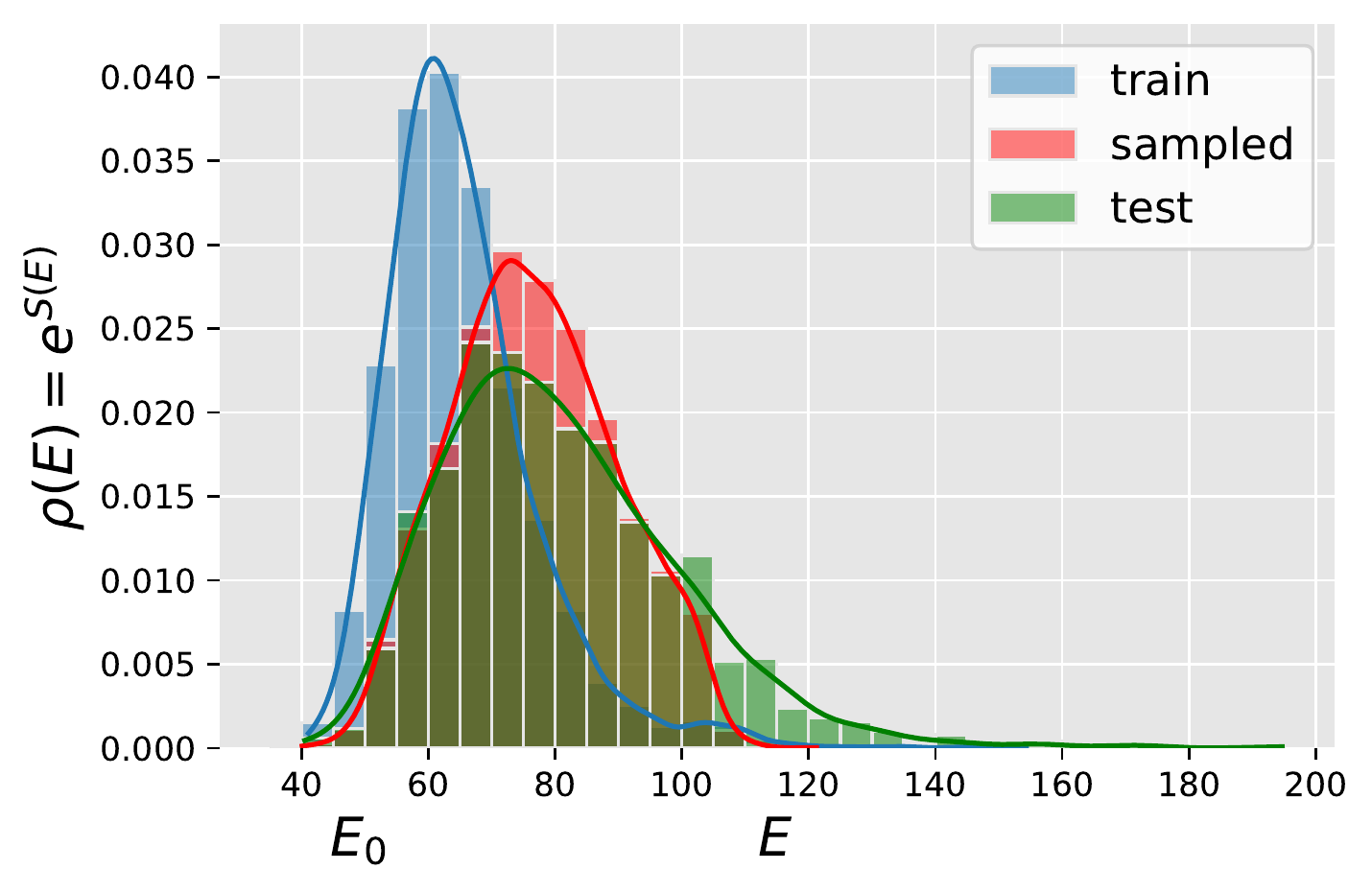}\\
\caption{Normalized distributions $\rho(E)dE$ of energies $E(x)$ for images $x$ from the train (blue), sampled (red), and test (green) sets correspondingly. Solid lines are kernel density estimations.}
\label{fig:rho}
\end{center}
\end{figure}

\begin{figure}
\begin{center}
\includegraphics[width=0.9\textwidth]{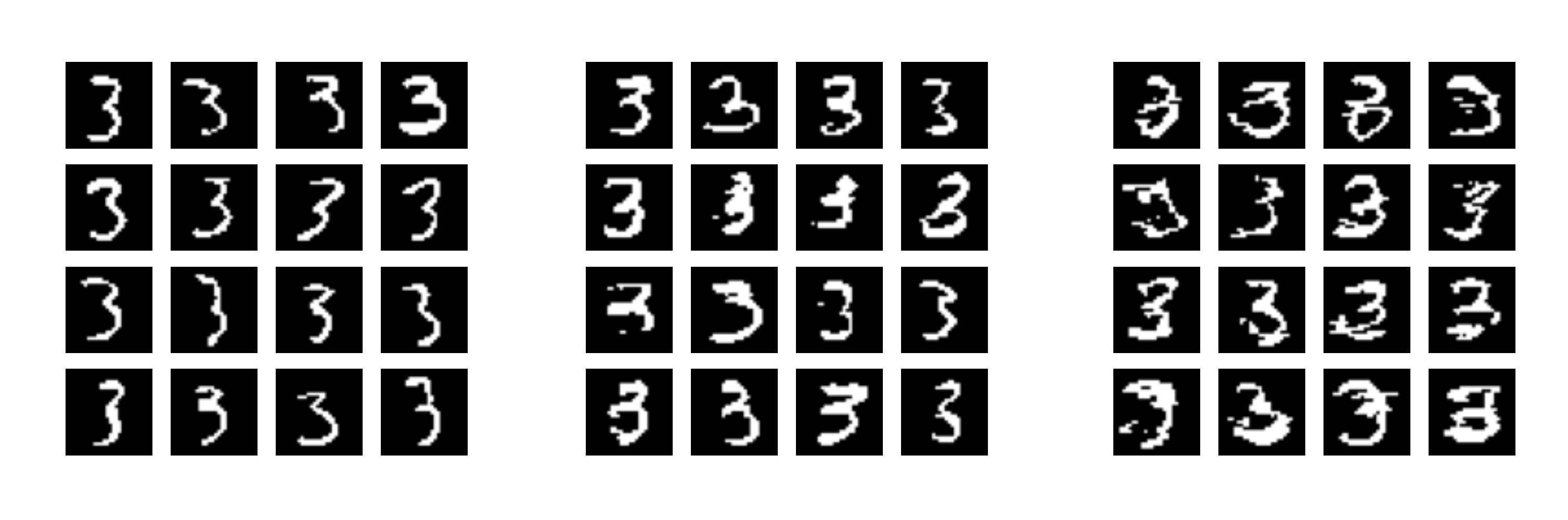}\\
\caption{Typical images sampled with help of $\Psi_3$ with three different ranges of  $E$, from left to right $E=60 \pm 5, 85 \pm 10, 125 \pm 10$.}
\label{fig:differentE}
\end{center}
\end{figure}

Next we discuss to what extent \eqref{P} defines  characteristic function of the full set of images of a given digit $i$.  We also address the question of the size 
$N_F$ of the full set  -- the global property of the full set which can not be deduces directly from the training dataset.
In what follows we focus on $i=3$ while results for other nine digits are qualitatively similar. 
First we would like to understand how many different images $x\in I$ there are with the given value of $|\Psi(x)|^2$  and what different values of $|\Psi(x)|^2$  represent. It would prove useful to use the language of statistical mechanics and think about different images $x\in I$ as possible microstates of some auxiliary physical system.\footnote{Familiarity with the basics of statistical mechanics are helpful but not necessary in what follows.} We define ``energy'' of  a microstate  $E(x):=-\ln|\Psi(x)|^2$ and entropy $S(E)$ such that $e^{S(E)} dE$ is the number of microstates with the energy between $E$ and $E+dE$.\footnote{As in statistical mechanics since microstates are discrete $e^S$ is a sum of delta-functions, which can be approximated by a smooth function when the total number of states $N_I$ is large, $N_I \gg 1$. In our case $N_I=2^{784}$ and this condition is well satisfied.} Then 
\bea
\int_0^\infty e^{S(E)} dE=N_I.
\eea
In these notations normalization condition \eqref{normalization} becomes unity of the ``partition function'' at unit temperature 
\bea
Z=\int_0^\infty e^{S(E)-E} dE=1.
\eea
We first discuss the distribution of energies $E(x)$ of the training set, which we denote $\rho_{\rm tr}(E) dE$, shown in blue in Fig.~\ref{fig:rho}. Minimization of loss function \eqref{lossf} is the minimization of  energy averaged over  $\rho_{\rm tr}$. The shape of $\rho_{\rm tr}$  is not robust and the mean value of the loss function $\langle E\rangle_{\rm tr}$ decreases with the increase of $D$.
What remains essentially the same is the energy of the lowest states, which we define by averaging $|\Psi(x)|^2$ over the training set
\bea
E_0=-\ln\langle e^{-E}\rangle_{\rm tr}. \label{E0}
\eea
The distribution $\rho_{\rm tr}$ should be compared with the distribution of energies for images sampled by the network itself. The probability of sampling an image $x$ is equal $|\Psi(x)|^2$ and therefore the distribution of $E$ for the sampled images is the Gibbs distribution  at unit temperature
\bea
\rho_{\rm sm}=e^{S(E)-E}. \label{Gibbs}
\eea
It is shown in red in Fig.~\ref{fig:rho}. %Naturally [need an explanation?], $\langle E \rangle_{\rm sm}$ is larger than $\langle E\rangle_{\rm tr}$ and the shape of \eqref{Gibbs} is also changing with $D$. At the same time energy of the lowest states is robust and matches \eqref{E0} with a good precision%
Naturally, $\langle E \rangle_{\rm sm}$ is larger than $\langle E\rangle_{\rm tr}$, i.e.~the value of the loss function averaged over the  set of sampled images is larger than the one for the training set. The shape of \eqref{Gibbs} is also changing with $D$. At the same time energy of the lowest states is robust and matches \eqref{E0} with a good precision

%\begin{figure}
%\begin{center}
%\includegraphics[width=0.4\textwidth]{qualityofsampling}\\
%\caption{Quality of sampling, as assessed by an auxiliary discriminant network (see Methods) as a function of $E(x)$.}
%\label{fig:qualityofsampling}
%\end{center}
%\end{figure}

\bea
E_0\approx -\ln\langle e^{-E}\rangle_{\rm sm}.
\eea
Moreover with good accuracy it is equal to  energy of the ground state 
$E_0\approx E_g \equiv \min_{x} E(x)$, were minimization can go over training set $x \in  T$ or the set of sampled images $x\in S$.
From $e^S=\rho_{\rm sm}e^E$ we find that $e^S$ is growing  rapidly with $E$, at least for energies around $E \sim \langle E\rangle_{\rm sm}$. In other words there are exponentially more images $x$ with larger values of $E(x)$. As  $E$  grows, quality  of sampled images deteriorates. 
%The result is shown in Fig.~\ref{fig:qualityofsampling}.
Qualitatively we can explain this as follows. Low-energy states $x$ with $E(x)\approx E_0$ are the high quality ``neat'' images of digit $3$, which will be recognized as such with almost $100\%$ confidence. Each neat image gives rise to many more ``corrupted'' images, which still can be recognized as $3$, albeit with a smaller confidence. These are the images with the larger values of  $E$. As the level of corruption grows, so is the total number of such images, and their typical  $E$ increases. This is demonstrated in Fig.~\ref{fig:differentE}, where we show typical sampled images with three  different values of $E$. 

From this discussion it is clear there is no sharp value of threshold $\epsilon$ to define the boundary of the full set. The size of the full set, the total number of images $x$ for which ${\cal P}(x)=1$, is dominated by the images with $E \approx-\ln\epsilon$,  which grows roughly as  $e^E \approx 1/\epsilon$. 

To unambiguously define the size of the full set,  we define the latter it to include only neat images of $3$, in which case the threshold, which we will call $\varepsilon$, can be taken very close to $E_0\approx E_{\rm g}$. 
%To avoid this ambiguity we will define the full set to include only neat images of $3$, in which case the threshold $\epsilon$ can be taken very close to $E_0\approx E_{\rm g}$. % 
We propose   
a way to fix $\varepsilon$ in Methods, but provided $\Delta E=\varepsilon-E_{\rm g}$ is small enough, at leading order the total size of the full set will be given by the exponent, while $\Delta E$ will control the subleading term, 
\bea
\label{NF}
N_F=\int_{E_{\rm g}}^\varepsilon e^S dE,\qquad \ln N_F\approx E_0+ \ln (\rho_{\rm sm}(E_0)\Delta E)\approx E_0.
\eea

\begin{figure}
\begin{center}
\includegraphics[width=0.5\textwidth]{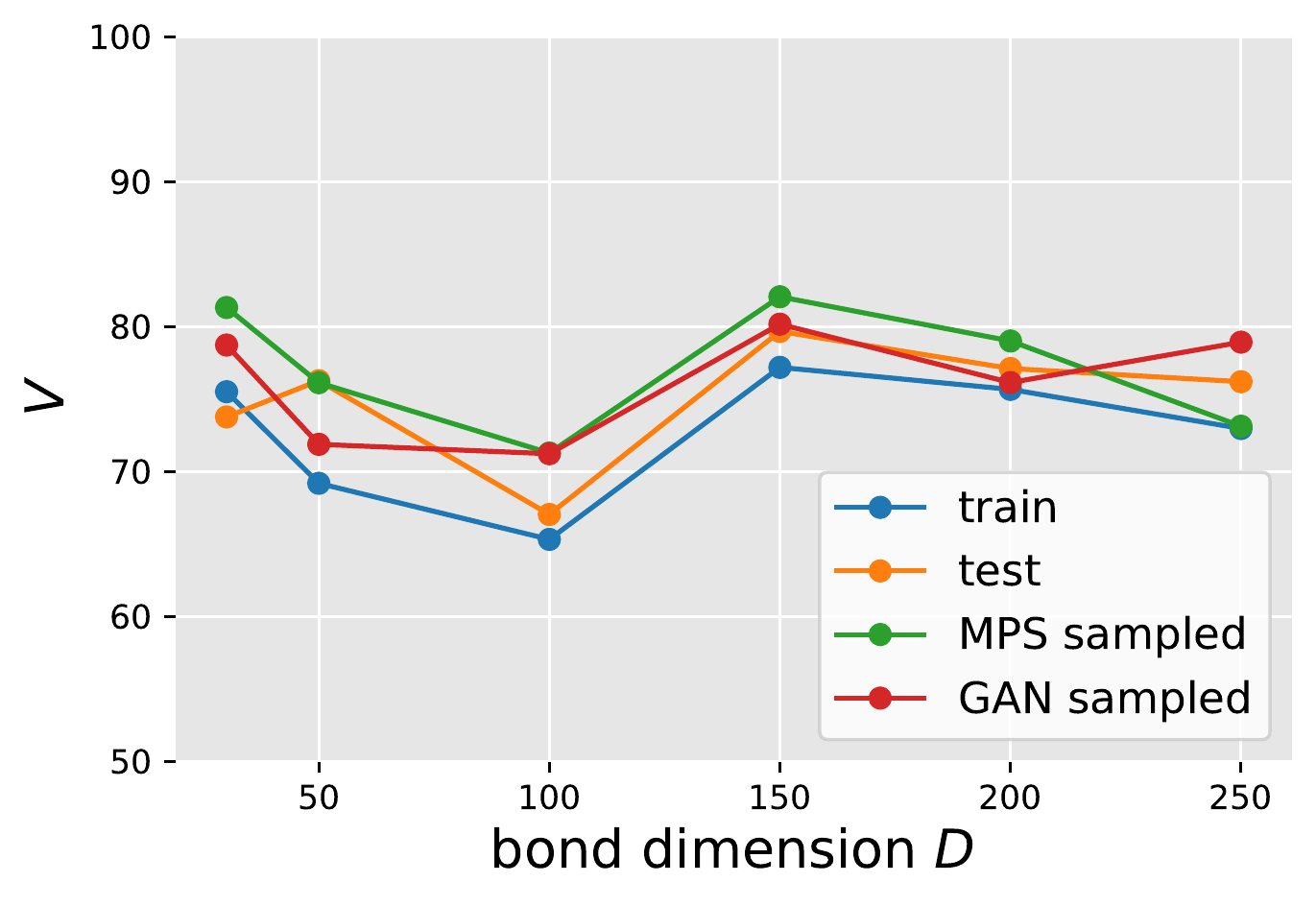}\\
\caption{The value of $E_0$ evaluated for different sets: training and test sets as well as sets images sampled with $\Psi$ itself and with help of an auxiliary GAN. All results are shown for trained
$\Psi_3(x)$ as a function of bond dimension $D$.}
\label{fig:E0_robust}
\end{center}
\end{figure}

The number  $N_F$ can be interpreted as both, the leading exponent controlling the size of the full set -- the total number of neat images of $3$, 
\bea
N_F\sim e^{E_0}\equiv 2^V,\qquad V\equiv E_0/\ln 2, \label{size}
\eea
%as well as the number of images $M$ which need to be sampled before there would be repetitions, i.e.~an image sampled twice.  The latter interpretation follows from  $M\rho_{\rm sm}=e^S$ understood as the equation on $E$  to minimize $M$. [do I need to explain this?]  An immediate question is how robust the latter interpretation is given that $\rho_{\rm sm}$ depends on the network architecture. We test it by sampling images with a trained network  with the bound dimension $D$ and evaluating $E(x)$ using another trained network with bond $\tilde{D}$. Alternatively, we sample images  using properly trained GAN. In both cases resulting $\rho_{\rm sm}$ have the same value of $E_0=-\ln \langle e^{-E} \rangle_{\rm sm}$ and therefore $M\sim e^{E_0}$. The comparison of different $\rho_{\rm sm}$ is shown in Fig.~\ref{}. [we probably need plot(s) showing robustness of $E_0$, namely $E_0$ evaluated using the training set and different sampled set and when applies for different $D$]%
as well as the number of images $M \approx N_F$ which need to be sampled before there would be repetitions, i.e.~an image sampled twice.  The latter interpretation follows from equating the total number of sampled images with given energy $E$,  $M\rho_{\rm sm}(E)$ and the total number of images $e^{S(E)}$ with this energy. Understood as the equation on $M$, it  yields $M=e^E$, where $E$ should be larger than $E_{\rm g}\approx E_0$. Minimization of $M$ over all possible value of $E$ readily gives \eqref{size}.
An immediate question is how robust the latter interpretation is, given that $\rho_{\rm sm}$ depends on the network architecture. We test it by sampling images with one trained network, with the bond dimension $D$, and evaluating $E(x)$ using another trained network, with  the bond dimension $D'$. Alternatively, we sample images  using properly trained GAN. In all cases resulting $\rho_{\rm sm}$ have  approximately the same value of $E_0=-\ln \langle e^{-E} \rangle_{\rm sm}$ and therefore $N_F\sim M\sim e^{E_0}$ remain the same. The comparison of  $E_0$ evaluated for $\Psi_3$ for different sets, training, test and sampled with help of $\Psi_3$ itself  and with an auxiliary GAN,  is shown in Fig.~\ref{fig:E0_robust}.

At this point we would like to conclude the energy of low lying states $E_0$ is a robust characteristic of the full set which defines its size, thus answering the question from the abstract. Here  the full set would be defined  to include only neat images of $3$ and similarly  for other digits.
The results for size $V$ defined  in  \eqref{size} for all ten digits  are shown in Table \ref{tab:thetable}. 

Yet a closer look reveals ${\cal P}(x)$ of a trained network is not quite the characteristic function of the full set we hoped it would be. Looking at the distribution $\rho_{\rm test}$ of energies $E(x)$ for the images from the test set, we immediately find many neat images of $3$ with the energies of order $\langle E\rangle_{\rm test}$, which  are significantly larger than $E_0$. This is clear from Fig.~\ref{fig:rho}, where $\rho_{\rm test}$ is shown in green.  This indicates not a conceptual flaw but certain deficiency in how our network was trained. We argue now,  for an idealized properly trained network typical values of $E(x)$ for images from both training and test sets should be around $E_0$. To confirm this we retrain our network by doubling the training set  using GAN-generated images. As expected, as the size of the training set $N_T$ increases both 
$\langle E\rangle_{\rm tr}$ and $\langle E\rangle_{\rm test}$ decrease, but  the value of $E_0=-\ln\langle e^{-E}\rangle$ defined with help of any set, train, test, or sampled, remains robust. The resulting picture is as follows. At  leading order the total size of the full set is given by \eqref{size} and is accessible by a network trained with help of MNIST, while the number of neat images of $3$ which our network misclassifies by assigning ${\cal P}(x)=0$ is substantially smaller than $e^{E_0}$.

\begin{table}
\begin{center}

\begin{tabular}{ |c|c|c|c|c|c|c|c|c|c|c| } 
 \hline
  $i$ & $V$ & $\langle d_{ab} \rangle$ &  $\Delta$ & $n$  & $\bar S$  \\ 
  \hline
 \bf{0} & $90 \pm 8$ & $133 \pm 3$ & 13  &  $138 \pm 33$
 & $4.4 \pm 0.2$ \\
 \bf{1}& $22 \pm 1$  & $60 \pm 25$ & 4 & $60 \pm 17$ & $3.6 \pm 0.1$\\ 
\bf{2} & $84 \pm 5$  & $133 \pm 27$ & 12 & $118 \pm 30$ & $3.9 \pm 0.2$\\
\bf{3} & $72 \pm 4$ & $121 \pm 29$ & 12 & $112 \pm 30$ & $4.0 \pm 0.1$\\
\bf{4} & $69 \pm 4$ & $109 \pm 26$& 11 & $96 \pm 26$ & $4. 1\pm 0.2$\\
\bf{5} & $77 \pm 4$ & $127 \pm 31$ & 11  & $102 \pm 30$ & $3.8 \pm 0.1$\\ 
\bf{6} & $71 \pm 4$ & $115 \pm 31$ & 11 & $109 \pm 30$ & $3.8 \pm 0.1$\\
\bf{7} & $53 \pm 2$ & $99 \pm 28$ & 9 & $91 \pm 24$ & $3.7 \pm 0.1$\\
\bf{8} & $92 \pm 4$ & $124 \pm 29$ & 13 & $120 \pm 31$ & $4.0 \pm 0.1$\\
\bf{9} & $60 \pm 5$ & $105 \pm 29$ & 9  & $97 \pm 26$ & $3.8 \pm 0.2$\\
 \hline

\end{tabular}
\caption{\label{tab:thetable} The table summarizes main properties of the full sets associated with the digits of black and white MNIST.  $V$, defined in \eqref{size}, controls  the size of the full set. $\langle d_{ab} \rangle$ is the  mean Hamming distance.  $\Delta$ is the full set effective  (fractal) dimension.  $n$ is an average number of black pixels for images of a given digit $i$.  $\bar{S}$ is the (averaged) entanglement entropy.}

\end{center}
\end{table}

\begin{figure}
\centering
\begin{minipage}{.5\textwidth}
  \centering
  \includegraphics[width=.9\textwidth]{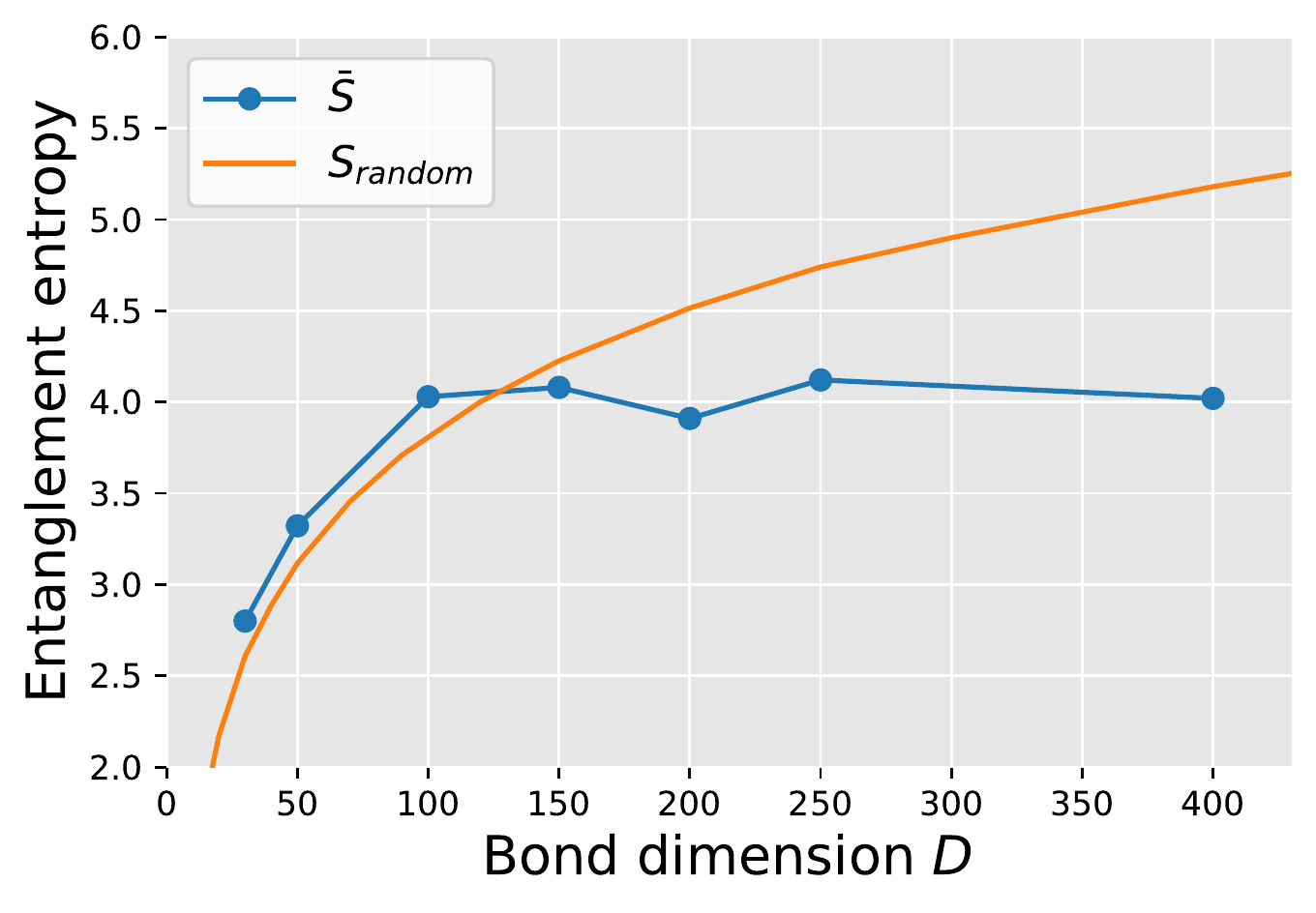}
%\caption{Average entanglement vs bond dim (definitely need to collect more data here)}
%  \label{fig:classifier}
\end{minipage}%
\begin{minipage}{.5\textwidth}
  \centering
 \includegraphics[width=.9\textwidth]{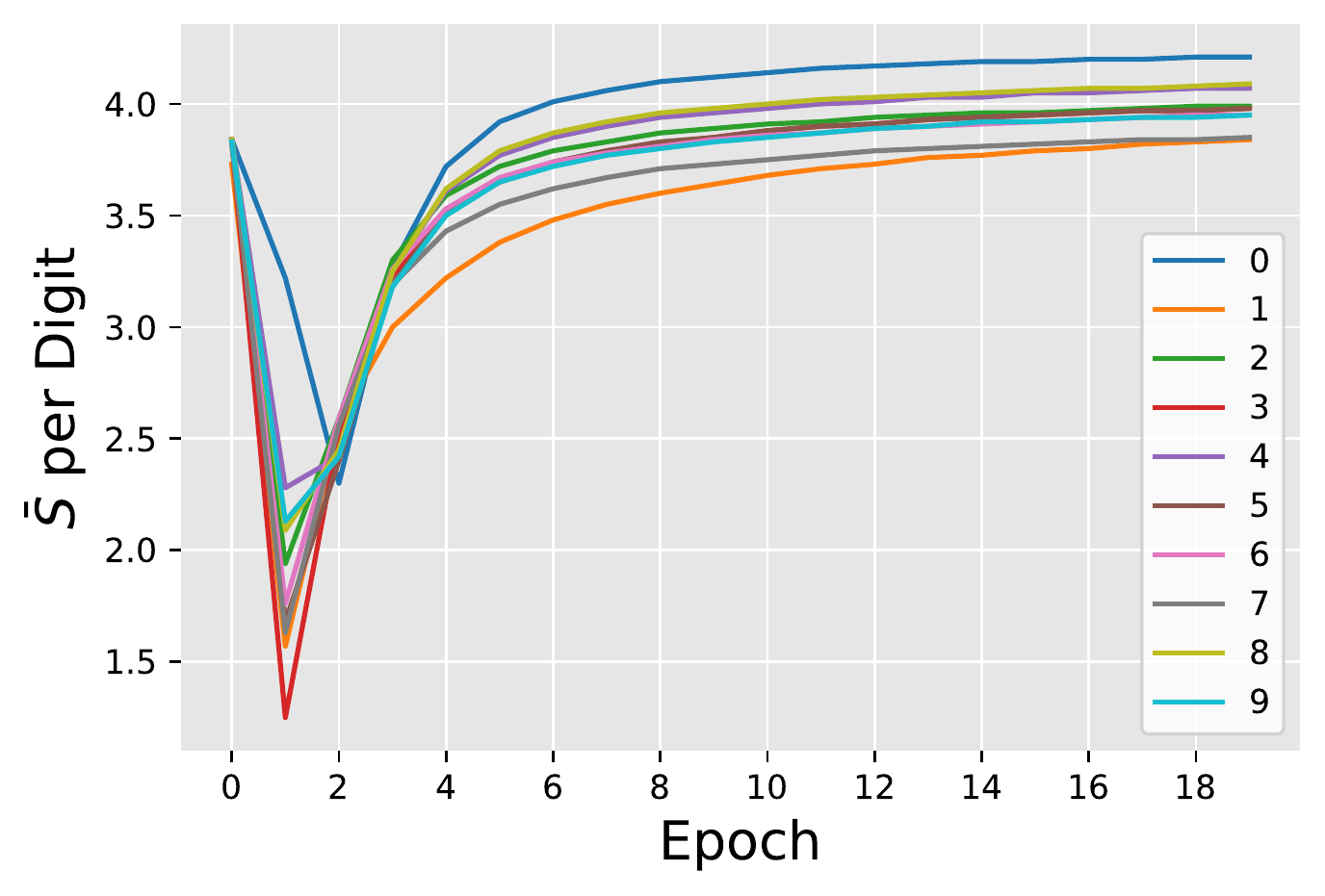}\\
%\caption{This is average entanglement (averaged over bipartitions in range from 200 to 600), while training to \textbf{sample}.}
%  \label{fig:sampler}
\end{minipage}
\caption{Left: Average entanglement $\bar{S}$ of $\Psi_3$ as a function of bond dimension.  Right: Average entanglement $\bar{S}$ for each $\Psi_i$ during training. }
\label{fig:EE}
\end{figure}

To further  characterize the full set geometrically, we evaluate its mean distance and  the effective dimension, defined with help of the Hamming distance. Individual images $x$ are binary strings and Hamming distance $d(x_1,x_2)$, defined as the number of distinct components of $x_1$ and $x_2$, provides a simple notion of distance between them. Clearly, Hamming distance is primitive in the sense  it does not reflect how similar or different the essence of  images would be to a human observer. Nevertheless the full set  understood as a subset of vertexes $x$ of a unit cube equipped with the Hamming distance, satisfying ${\cal P}(x)=1$
exhibits well-defined coarse grained geometric properties. The mean distance $d(x_1,x_2)$ between two random images of digit $i$, taken either from train/test or sampled sets is substantially smaller than two completely random images with the same mean value of black and white pixels. This is demonstrated in Table \ref{tab:thetable} where we show results for all ten digits.  

\begin{figure}
\begin{center}
\includegraphics[width=0.6\textwidth]{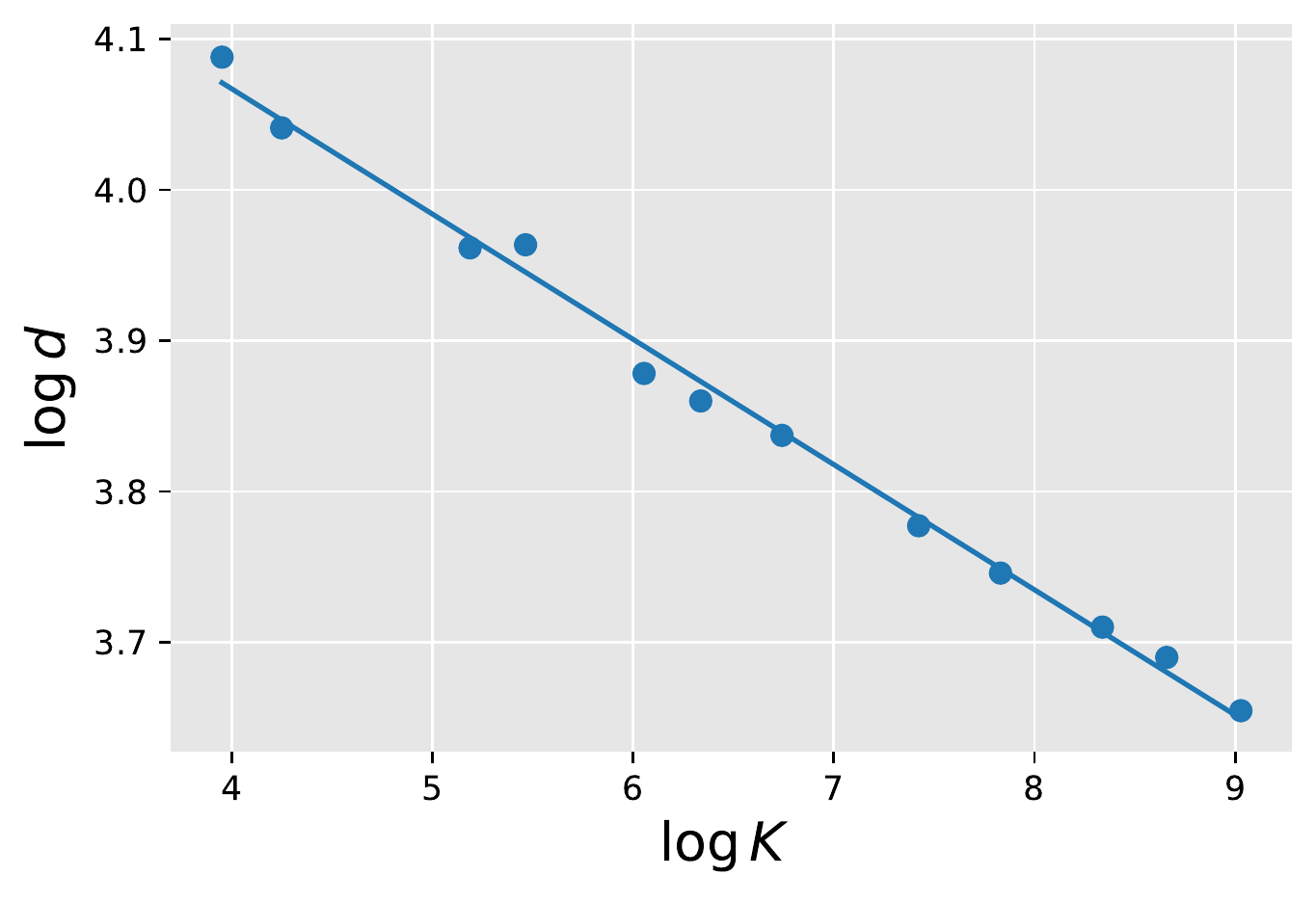}\\
\caption{Log-log plot for $d$ as a function of number of samples  $K$. Images are sampled with  $\Psi_3$ with the bond dimension $D = 100$. Only  images with $E(x) \approx E_0$ are included.}
\label{fig:fractallog}
\end{center}
\end{figure}

To evaluate the full set effective dimension, we use the standard approach of \cite{Hein2005intrinsic,  Facco2017dim}.  For a set of images $x_a$ we define  minimal distance 
\bea
d_{\rm min}(x_a)=\min_b d(x_a,x_b)
\eea
and then study how mean value $d=\langle d_{\rm min}(x_a)\rangle$, averaged over the set of $x_a$, scales with the set size $K$
\bea
d \propto K^{-\Delta}.
\eea 
Here $\Delta$ is the effective (or fractal)  dimension. 
Linear fit of $d(K)$ in log-log coordinates is shown in Fig.~\ref{fig:fractallog}.
We focus on neat images of digits $i$. Therefore, we consider a subset of train/test and sampled sets for which $E(x_a)$ is close to $E_0$. In both cases we obtain similar results, indicating full sets of  digits have well-defined effective dimensions. The results are shown in Table \ref{tab:thetable}. We would like to note,  unlike the size of the full set, which is a global property requiring knowledge of the whole full set, mean  distance and the effective dimensions can be inferred  directly from the train/test set. 
Our results for $\Delta$ are compatible with previous studies of the effective dimension \cite{Hein2005intrinsic,  Facco2017dim},
which used Euclidean distance in conjunction with the original MNIST. This  suggests rendering images black and white does not drastically affect geometries properties of the full set.

\begin{figure}
\begin{center}
\includegraphics[width=0.6\textwidth]{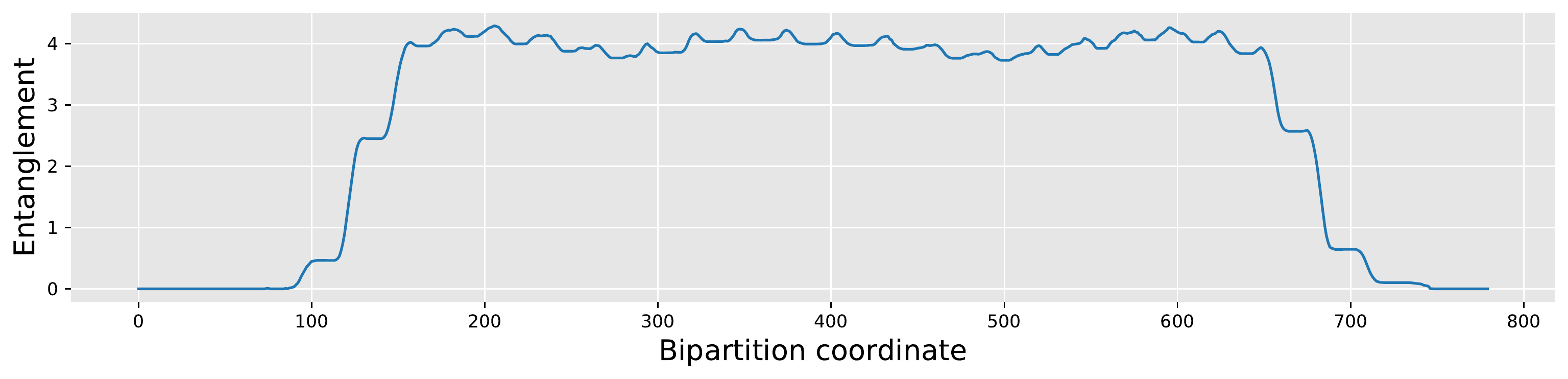}\\
\caption{Page curve for $\Psi_3$ with the bond dimension $D = 100$.}
\label{fig:Page}
\end{center}
\end{figure}

%$\Psi$, that characterizes the data itself, 
As an application of our architecture, we evaluate  entanglement entropy (EE), by interpreting $\Psi$ as a quantum-mechanical wavefunction. For a tensor network its maximal EE determines its expressiveness. 
In the context of quantum physics EE is a central quantity which measures the amount of information shared between different parts of the system \cite{RevModPhys.82.277}. In particular it rigorously bounds classical mutual information associated with a bi-partition \cite{Um_2012}
\bea
I(A,B)\leq S_{AB}.
\eea 
Popularity of EE transcended physics into machine learning with several different groups recently discussing it in the context of tensor-based architectures \cite{pestun2017tensor,  pestun2017language, levine2017deep, levine19quantum, martyn2020entanglement}. For a MPS architecture it is natural to discuss EE of bi-partitions, $S_k$, where all $n=784$ pixels are split into two groups of $k$ and $n-k$ pixels correspondingly, see Methods. 
Resulting {\it Page curve}\footnote{The name comes from the pioneering work evaluating EE of random states \cite{page1993average}.}   is shown in Fig.~\ref{fig:Page}.
We take EE averaged along the ``plateau" region of  bi-partitions with $k$ ranging between $200$ and $600$, for which $S_k$ is approximately the same. This corresponds to splitting  image horizontally into two  halves.
Averaged EE,  which we denote by $\bar S$,  slowly grows with epoch, which is expected: as the tensor network tries to better fit training data it needs more expressiveness associated with larger entanglement. Notably, after a few epochs $\bar S$ become essentially independent of the initial seed. We stop training $\Psi$  when   it exhibits  maximal quality of discrimination/classification. Averaged $\bar S$ for such $\Psi$ as a function of  bond dimension is shown in the left panel of Fig.~\ref{fig:EE}. It quickly grows for small $D$ and becomes approximately constant for $D\gtrsim 100$. 
Robust  independence  of $\bar S$ on  $D$ and the initial seed is a  further confirmation EE of a trained network is a characteristic of the data itself, not of the network architecture.
This is the crucial difference between our work and previous studies of the  EE  in the context of tensor network algorithms. Our value of $\bar S$  should be contrasted with the maximal EE $S_{\rm max}=2\ln D$ for a tensor network with the bond dimension $D$, and for a random network with all MPS tensors drawn from the unitary ensemble, $S_{\rm random} \simeq \ln D$, see Fig.~\ref{fig:EE}.

Relatively small value of $\bar S$, and hence of mutual information between upper and lower halves of the  images for all  ten digits indicates there is a small number of ways to continue the image of a given digit, if a  half of it is known. Schematically, this means there is a finite number of styles to handwrite  any given digit $i$: once an upper part of the image is given, it fixes the digit itself and its style within a range of a few possibilities. 
%Say, an upper part of $3$ shown in Fig.~\ref{??} indicates it is likely $3$ or $7$, and style of each digit is approximately fixed.  
This interpretation is corroborated by a positive correlation between the value of the entanglement, shown in Table \ref{tab:thetable}, and  sizes of the full sets $N_F$, see eq. (\ref{size}). The logic here being that larger number of styles will be reflected by larger value of $N_F$.

%[In fact EE gives a reasonable estimate for mutual information as is shown in Fig.~\ref{Vidal,Cirack}.]
The EE provides an upper bound on mutual information, an important information-theoretic properties characterizing the data.  Recently mutual information and the entanglement entropy of data in the training set have been studied in  \cite{martyn2020entanglement,  deng17ent,  convy2021mutual,  lu2021tensor}. Our work provides an alternative conceptually better way to evaluate it, as it is based on the full set, rather than the training set alone.

%[EE about zero]

%\begin{figure}
%\begin{center}
%\includegraphics[width=0.6\textwidth]{random}\\
%\caption{Random isometry entanglement vs maximum entanglement acheived by our TN during training}
%\end{center}
%\end{figure}

\section*{Discussion}
In this paper we introduced a tensor network architecture to learn the wavefunction of data $\Psi$. We introduced the notion  ``the full set of data'' -- the collection of all hypothetical data exhibiting the same underlying pattern. Our tensor network 
provides a practical way to learn and subsequently characterize  the full set by defining its indicator function ${\cal P}(x)$, see eq.~\eqref{P}. We have trained $\Psi$ using black and white version of MNIST and demonstrated its  core properties  are independent of the network parameters, which confirms they are characteristics of the data itself. Using $\Psi$ we have estimated the sizes of the full sets of individual digits, i.e.~the total number of black and white MNIST-style images depicting a particular digit $i$. The results are shown in Table \ref{tab:thetable}. To further illustrate utility of $\Psi$ as a vehicle to study the data itself, we have  calculated entanglement entropy, which upper bounds mutual information, associated with splitting  images into two parts. The results are also shown in Table \ref{tab:thetable}.  The entanglement entropy/mutual information of MNIST images is small, which indicates relatively small number of different styles of handwriting. 

The full set  is a concrete, learnable cousin of an abstraction called the manifold of data  in the series of recent papers \cite{kaplan2020scaling,  sharma2020neural,  bahri2021explaining}.
The manifold of data, by definition, requires  idealized ``infinite data limit'' when the training set grows indefinitely. This is in contrast with our approach suited for practically available datasets. We have seen  in the case of MNIST digits, certain images from the train/test set fall outside of the full set defined via indicator function \eqref{P}. We argued, this problem goes away as the size of the training set grows. In this case the distributions $\rho_{\rm tr}$, $\rho_{\rm sm}$, and $\rho_{\rm test}$ have smaller support and eventually, in the infinite data limit, collapse  to a narrow vicinity of $E_0$. This is the limit in which the full set, which would striclly include all images of a given digit, would become the manifold of data  of refs.~\cite{kaplan2020scaling,  sharma2020neural,  bahri2021explaining}.\footnote{ In our case the space of images $I$ is discrete, while the data  discussed in \cite{sharma2020neural} is parametrized by vectors in ${\mathbb R}^n$. Hence the full set in the infinite data limit will became a discrete version of the manifold of data.} The rate with which  the value of the loss function $\langle E\rangle_{\rm tr}-E_{\rm g}$ and  $\langle E\rangle_{\rm test}-E_{\rm g}$ decreases with $N_T$ -- the size of the training set, as well its dependence on  bond dimension $D$, should presumably follow the universal scaling laws outlined in \cite{kaplan2020scaling,  sharma2020neural,  bahri2021explaining}. To verify that would be an interesting problem.

One of the most important open questions of machine learning is to understand  why certain datasets admit generalizations, as is the case of virtually any visual dataset exhibiting a pattern recognizable by a human observer. It would be a substantial step forward to quantitatively characterize this phenomenon by explaining why generalization is possible. The studies of mutual information/entanglement entropy, initially motivated by a more narrow question of gauging suitability of tensor network-based architectures, is a first step in this direction. It is clear though, a much more comprehensive characteristic of data is necessary to understand if it admits generalization  and the best way to achieve it.  Our network is a  novel tool to characterize the full set of data globally through the  boolean function $\cal P$, in the present case defined on a unit cube of the dimension $n=784$.  We surmise that the boolean function complexity \cite{Jukna2012} could be the right language to characterize possible efficiency of generalization in the quantitative terms. Furthermore, drawing from the connection with quantum mechanics, we believe {\it circuit complexity} of $\Psi$ interpreted as a quantum-mechanical wave-function could be an important characteristic of patterns, underlying the  initial dataset.

%[novel tool to characterize full set of data globally through boolean function $\cal P$; complexity of data = circuit complexity; goes beyond entanglement entropy]
%
%[collapse of $\rho_{\rm tr}$, $\rho_{\rm sm}$, and $\rho_{\rm test}$ to a narrow vicinity of $E_0$, scaling, connection to papers by Kaplan]
%
%[open question - what makes data simple and ``learnable''?]
%
%[From the mathematical point of view \eqref{P} is a Boolean function on a unit cube of dimension $n=784$ \dots]
%
%[Suppmental Material/Appendix -- discussion of degeneracy of $|\Psi(x)|^2$ via stat mech?]

\section*{Methods}
\subsection*{MPS training procedure}

\begin{figure}
\begin{center}
\includegraphics[width=0.6\textwidth]{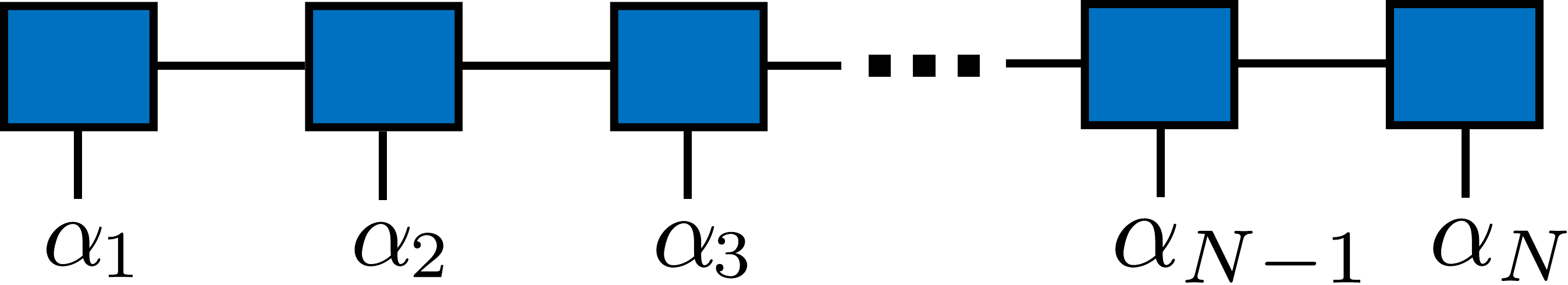}\\
\caption{Matrix Product State graph representation. Blue boxes represent tensors (of rank 3) with $\alpha_i$ external indices, associated with pixel representation. Bond dimension $D$ controls the maximum of other two dimensions of the tensors, depicted by horizontal lines.}
\label{fig:mps}
\end{center}
\end{figure}

%In this section we describe the details of training procedure. 
We train a separate MPS $\ket{\Psi_i}$ for each label $i$ in the dataset ($i \in 0..9$ in case of MNIST).  Our training procedure is similar  to \cite{Han18generative,  Cheng19generative},  with the key difference being the use of  tangent-space gradient optimization of \cite{Sun20TSGO}.  We start by mapping training  samples $x$ to  product states $\ket{x} = \prod_{\otimes \alpha} \ket{x^{\alpha}}$ by representing each black or white pixel $x^{\alpha}$ with ``down'' $\ket{0}$ and ``up'' $\ket{1}$ states. With this representation of data samples we can define the probability of a given sample in accordance with Born's rules as    
\bea
p(x) = \frac{|\bra{x}\ket{\Psi}|^2}{\bra{\Psi}\ket{\Psi}}.
\eea
We train out network by minimizing values of the probability distribution $p(x)$ averaged over the training set $T$, while keeping normalization condition (\ref{normalization}) for the wave function $\bra{\Psi}\ket{\Psi} = 1$. Namely, we minimize the negative-log likelihood (NLL) loss function $L$ during training,
\bea
\label{loss}
L = - \frac{1}{N_T} \sum_{x \in T} \log p(x).
\eea

The gradient of the NLL loss function can be found analytically, 
\bea
\frac{\partial 	L}{\partial \ket{\Psi}} = 2 \ket{\Psi} - \frac{2}{N_T} \sum_{x \in T} \frac{\ket{x}}{\bra{x}\ket{\Psi}}.
\eea 

In practice we do not evaluate gradient with respect to $\ket{\Psi}$,  instead we update each tensor of MPS independently via DMRG  \cite{Han18generative,  Cheng19generative} method with a two-site update. 

Gradient descent is carried out by TSGO \cite{Sun20TSGO} with rotation angle (learning rate) $\eta = \pi/36$,  which showed better and faster convergence compared to Adam or SGD.

\textbf{Sampling.} Sampling from a trained $\Psi$ is carried via Born rule.  Starting from the first pixel, which corresponds to the state $\ket{x^1}$, one samples it with  marginal probability $p(x^1) = \frac{|\bra{x^1}\ket{\Psi}|^2}{\bra{\Psi}\ket{\Psi}}$.  Here $\bra{x^1}\ket{\Psi}$ is a ``state'', i.e.~a tensor with $N-1$ indexes $\alpha_2,\alpha_2,\dots, \alpha_N$. 
The subsequent pixel is sampled conditionally: $p(x^{2} | x^{1}) = p(x^{1} \cup x^{2})/p(x^1)$.  To effectively calculate the probabilities one needs to keep the MPS in the right canonical form \cite{Han18generative}. Typical examples of sampled images are shown in Fig.~\ref{fig:differentE}.

\textbf{Quality of sampling.} To assess quality of sampling we trained an auxiliary CNN to classify MNIST-like images on QMNIST \cite{qmnist-2019}, a dataset which extends MNIST with additional 50K images, which allowed us to train CNN on images never seen by our MPS. We sampled images with MPS and passed them through CNN which returned a probability of sampled image of, say, three to be classified as three. This probability we interpret as the quality of sampling, which is shown in the left panels of Fig.~\ref{fig:training} and Fig.~\ref{fig:quality}.

\textbf{Classification.} To classify images we take ten MPS $\Psi_i$ trained to minimize the loss function (\ref{loss})  for each digit $i$ separately. To predict label for an image $x$ we calculate $\text{arg}\,\max\limits_{i}\,  |\Psi_i(x)|^2$. Accuracy of such classification peaked at around $96\%$ in our simulations, which is not as high as contemporary supervised NN architectures, but on par with common unsupervised methods. 

\textbf{Discrimination.} Discrimination quality is the ability of a NN to distinguish images of, say, threes from any other images. In our case, due to normalization condition (\ref{normalization}) the wave function $\Psi_3$  tends to decrease outside of the set of threes. In our setup $ |\Psi(x)|^2 \geq  \epsilon$ indicates that $x$ is contained in the set of threes and $x$ is outside of the set of threes if $ |\Psi(x)|^2 <  \epsilon$. Here we propose a simple method to fix $\epsilon$ and estimate discrimination quality. While assessment of discrimination quality in principle requires to test MPS against the whole ``space of images" $I$, in practice, most of this space is irrelevant. Our model does not confuse threes with random images. The hardest examples to discriminate  are pictures which are structurally similar to 3s, i.e.~those depicting other symbols. That is why we test $\Psi_i$ on other digits on MNIST. We split training test into two parts. One part is used to train the model, while the other is used to fix $\epsilon$ such that the balanced accuracy on the test set is maximized.  The resulting quality of discrimination is shown on the right panel of  Fig.~\ref{fig:training}.

\textbf{Fixing $\varepsilon$.} The value of $\varepsilon$, introduced around eq.~\eqref{NF} should be fixed to  separate ``neat" images of digit $i$ from the ``corrupted" ones. There is no precise definition of ``neat" or ``corrupted". However, we can try to estimate $\varepsilon$ as follows. Firstly, we use trained $\Psi_i$ to sample the images of $i$ and keep only those with the value of $E   \approx  E_0$. With the help of auxiliary CNN, we estimate sampling quality of each sample and calculate mean and variance of the sampling quality distribution. Then, we start looking into sampled images with larger values of $E > E_0$. For the given $E$ we estimate mean and variance of the sampling quality distribution. Thus for each $E$ we have mean value and variance. Now we are looking for $E = E_*$  which is large enough such that corresponding sampling quality distribution is sufficiently different from the one for $E = E_0$, namely corresponding mean values are separated by at least the standard deviation. This value of $E_*$ is taken to be $\varepsilon = E_*$. We have checked numerically that $\varepsilon$ obtained in this way yields sub-leading contribution to $N_F$, see eq.~(\ref{size}).

\textbf{Further comments.} We trained an auxiliary Deep Convolutional GAN to independently sample additional MNIST-like images. With this samples we doubled the amount of images and retrained MPS on extended data. As the result the value of mean $\langle E \rangle$ averaged over test set decreased by $4\%.$  Additionally we used GAN samples to verify our estimation of effective  (fractal) dimension for each digit. The results are in agreement with values of $\Delta$ from the Table \ref{tab:thetable}.

\subsection*{Entanglement Entropy} 
Entanglement Entropy of a bi-partition which splits a quantum spin-chain into two parts $A$ and $B$ consisting  of $k$ ``left'' and $n-k$ ``right'' spins correspondingly is defined as 
\bea
S(\rho_A) = - \Tr(\rho_A \log \rho_A) =  - \Tr(\rho_B \log \rho_B) = S(\rho_B), 
\eea
where $\rho_A = \Tr_B (\rho_{AB})$ and $\rho_B = \Tr_A (\rho_{AB})$ are reduced density matrices for each partition and $\rho_{AB} = \ket{\Psi} \bra{\Psi}$ for a pure state.

For any pure state the entanglement entropy can be expressed using the singular values of the Schmidt decomposition of the state,
\bea
\ket{\Psi} = \sum_{i} \lambda_i \ket{u_i}_A \otimes \ket{v_i}_B, \label{Schmidt}
\eea
where $\ket{u_i}_A$ and $\ket{v_i}_B$ are orthonormal states of the subsystems A and B separated . The entanglement entropy reduces to 
\bea
S(\rho_A) = S(\rho_B) = - \sum_i |\lambda_i|^2 \log (|\lambda_i|^2).
\eea
%The advantage of the MPS is that it can be brought to the canonical form in such a way that the leftmost spin of the %subsystem $B$ becomes the orthogonality center. \textcolor{red}{This index $\alpha$ exactly corresponds to the %Schmidt decomposition \cite{Vidal2003}.}
The advantage of the MPS is that we can bring network to the form with the orthogonality center to be the 
bond between the subsystems $A$ and $B$. Then Schmidt decomposition \eqref{Schmidt} becomes singular value decomposition of the product of two MPS tensors from  the nodes adjacent 
to the orthogonality center bond \cite{Vidal2003}.

%\bea
%S_{AB}=-\Tr(\rho \ln \rho),\qquad \rho_{\alpha_1\dots \alpha_k \alpha'_1\dots \alpha'_k}=\sum_{\beta}\dots  \%Psi_{\alpha_1,\dots,\alpha_k \beta_{k+1}\dots \beta_n}\Psi_{\alpha'_1,\dots,\alpha'_k \beta_{k+1}\dots \beta_n}.
%\eea

%The definition seems to break symmetry between $A$ and $B$ by contracting ``$B$'' indexes in the definition of $\rho%$, but the value of $S$ is independent of this chose. 
%In case of MPS and the bi-partition which splits it into two connected parts EE is very simple to calculate \dots 

%
%\section*{Data availability}
%All simulations were carried out using PyTorch machine learning library[??]. Original simulation results are available from the corresponding authors on a reasonable request.
%\section*{Code availability}
%\section*{References}
\section*{Acknowledgements}
We would like to thank  Vassilis Anagiannis,  Miranda Chen, and especially Vasily Pestun for collaboration at the early stages of this project. We also acknowledge helpful discussions with Dan Roberts, Ivan Oseledets, Miles Stoudenmire and David Schwab. 

%\section*{Author contributions}
%\section*{Competing interests}
%The authors declare no competing interests.
%
%
%
%
%
%
%
%
%
%\section*{References.}
%
%{\label{190269}}
%
%These may only contain citations and should list only one publication
%with each number. Include the title of the cited article or dataset.
%Acknowledgements (optional). Keep acknowledgements brief and do not
%include thanks to anonymous referees or editors, or effusive comments.
%Grant or contribution numbers may be acknowledged. Author contributions.
%You must include a statement that specifies the individual contributions
%of each co-author. For example: ``A.P.M. `contributed' Y and Z; B.T.R.
%`contributed' Y,'' etc. See our
%\href{http://www.nature.com/authors/policies/authorship.html}{authorship
%policies} for more details.
%

%\subsection*{Competing financial
%interests.}
%
%{\label{201484}}
%
%Submission of a competing financial interests statement is required for
%all content of the journal.
%
%\subsection*{Materials \&
%Correspondence.}
%
%{\label{309691}}
%
%Indicate the author(s) to whom correspondence and material requests
%should be addressed.
%
%\subsection*{Tables.}
%
%{\label{129844}}
%
%Each table should be submitted as a word document and accompanied by a
%short title sentence describing what the table shows. Further details
%can be included as footnotes to the table.
%
%\subsection*{Figures}
%
%{\label{530077}}
%
%High-resolution image files are not required at initial submission, but
%please ensure that images are of sufficient resolution for referees to
%properly assess the data.
%
%~

%\selectlanguage{english}
%\FloatBarrier

\bibliographystyle{unsrt}
\bibliography{TN}

\end{document}